\documentclass[graybox]{svmult}
\usepackage{mathptmx}       
\usepackage{helvet}         
\usepackage{courier}        
\usepackage{makeidx}         
\usepackage{graphicx}        
\usepackage[bottom]{footmisc}
\makeindex             
\begin{document}

\title{Physics and Complexity: a brief spin glass perspective}
\author{David Sherrington}
\institute{David Sherrington \at Rudolf Peierls Centre for Theoretical Physics,  1 Keble Rd., Oxford OX1 3NP, 
UK, and Santa Fe Institute, 1399 Hyde Park Rd., Santa Fe, NM 87501, USA.
\email{D.Sherrington1@physics.ox.ac.uk}}
%
%

\maketitle

7 Jan 2012
\\

\abstract{Complex macroscopic behaviour can arise in many-body systems
with only very simple elements as a consequence of the combination of competition and inhomogeneity.
This paper attempts to illustrate how statistical physics has driven this recognition, has contributed 
new insights and methodologies
of wide application influencing many fields of science, and has been stimulated in return.}

\section{Introduction}

Many body systems of even very simple microscopic constituents with very simple interaction rules can show 
novel emergence in their macroscopic behaviour. When the interactions (and any constraints) are also mutually 
incompatible (frustrated) and there is macroscopically relevant quenched disorder,  then the emergent 
macroscopic behaviour can be complex (in ways to be discussed) and not simply anticipatable. Recent years have 
seen major advances in understanding such behaviour, in recognizing conceptual ubiquities across many 
apparently different systems and in forging, transferring and applying new methodologies. Statistical physics 
has played a major part in driving and developing the subject and in providing new methods to study and 
quantify it. This paper is intended to provide a brief broadbrush introduction.

A key part of these developments has been the combination of minimalist modelling,  development of new concepts 
and techniques, and fruitful transfers of the knowledge between different systems. Here we shall concentrate on 
a simple paradigmic model, demonstrate its ubiquitousness among several often very different systems, problems 
and contexts, and introduce some of the useful concepts that have arisen.

\section{The Dean's Problem and Spin Glasses}
\label{sec:SKDean}

The genesis for the explosion of interest and activity in complexity within the physics community was in an 
attempt to understand a group of magnetic alloys known as spin glasses\footnote{Spin glasses were originally 
observed as magnetic alloys with unusual non-periodic spin ordering. They were also later recognized as having 
many other fascinating glassy properties}\cite{MPV}. But here we shall start with a problem that requires no 
physics to appreciate, the Dean's problem \cite{Clay}. 

A College Dean is faced with the task of distributing $N$ students between two 
dormitories as amicably as possible but given that some pairs of students prefer
to be in the same dorm while other pairs want to be separated. If any odd number of students have an odd number 
of 
antagonistic pairwise preferences then their preferences cannot all be satisfied simultaneously. This is an 
example of  frustration. 
The Dean's Problem can can be modelled as a mathematical optimization problem by defining a cost function 
{$\it{H}$} that is to be minimiized:
\begin{equation}
H=-\sum_{(ij)}J_{ij}\sigma_{i}\sigma_{j}; \sigma =\pm 1
\label{eq:SK}
\end{equation}
where the $i,j$ label students, ${\sigma=\pm 1}$ indicates 
dorm A/B
and the  $\{J_{ij}={J_{ji}}\}$ denote the sign and magnitude of the inter-student pair preferences (+ = 
prefer).
We shall further concentrate on the situation where the $\{J_{ij}\}$ are chosen randomly and independently from 
a single (intensive) distribution $\mathsf{P}(J)$ of zero mean\footnote{This restriction is not essential but 
represents the potentially hardest case.}, the random Dean's Problem. The number of combinations of possible 
choices grows exponentially in $N$ (as $2^{N}$). There is also, in general, no simple local iterative mode of 
solution. Hence, in general, when $N$ becomes large the Dean's problem becomes very hard,  in the language of 
computer science, NP-complete \cite{Garey}. 

In fact, the cost function of the random Dean's Problem was already introduced in 1975 as a potentially soluble 
model for a spin glass; there it is known as the Sherrington-Kirkpatrick (SK) model \cite{SK}. In this model 
$H$ is the Hamiltonian (or energy function), the ${i,j}$ label spins, the $\sigma$ their orientation (up/down) 
and the $\{J\}$ are the exchange interactions between pairs of spins. 

In the latter case one was naturally interested in the effects of temperature and of phase transitions as it is 
varied. In the standard procedure of Gibbsian statistical mechanics, in thermal equilibrium the probability of 
a microstate $\{\sigma\}$ is given by 
\begin{equation}
\mathcal{P}(\{\sigma\})=Z^{-1}\exp[-H_{\{J_{ij}\}}(\{\sigma\})/ T]
\label{eq:P(s)}
\end{equation}
where Z is the partition function
\begin{equation}
Z=\sum_{\{\sigma\}} \exp[-H_{\{J_{ij}\}}(\{\sigma\})/ T];
\label{eq:Partition_function}
\end{equation}
the subscript $\{J_{ij}\}$ has been added to $H$ to make explicit that it  is for the particular instance of 
the (random) choice of $\{J_{ij}\}$. Again in the spirit of statistical physics one may usefully consider 
typical physical properties over realizations of the quenched disorder, obtainable by averaging them over those 
choices \footnote{This is in contrast with traditional computer science which has been more concerned with 
worst instances.}. 

Solving the SK model has been a great challenge and has led to new and subtle mathematical techniques and 
theoretical conceptualizations, backed by new computer simulational methodologies and experimentation, the 
detailed discussion of which is beyond the scope of this brief report. However a brief sketch will be given of 
some of the conceptual deductions.  

Let us start pictorially. A cartoon of the situation is that of a hierarchically rugged landscape to describe 
the energy/cost as a function of position in the space of microscopic coordinates and such that for any local 
perturbations of the microscopic state that allow only downhill moves the system rapidly gets stuck and it is  
impossible to iterate to the true minimum or even a state close to it. Adding temperature allows also uphill 
moves with a probability related to $\exp [- {\delta H} /T]$ where $\delta H$ is the energy change. But still 
for $T < T_g$ the system has this glassy hindrance to equilibration, a non-ergodicity that shows up, for 
example, in differences in response functions measured with different historical protocols.

Theoretical studies of the SK model have given this picture substance, clarification and quantification, partly 
by introduction of new concepts beyond those of conventional statistical physics, especially through the work 
of Giorgio Parisi \cite{MPV, Parisi2005}. 

Let us assume that, at any temperature of 
interest, our system has possibly several essentially separate
macrostates, which we label by indices $\{S\}$. A useful measure of 
similarity of two microstates 
$S,S'$ is given by their `overlap', defined as 
\begin{equation}
q_{SS'}=N^{-1}\sum_{i} \langle \sigma_i \rangle_{S} 
\langle \sigma_i \rangle_{S'}.
\label{eq:overlap}
\end{equation}
where $\langle \sigma_i \rangle_{S}$ measures the thermal average of $\sigma_i$
in macrostate $S$.

The distribution of overlaps is given by
\begin{equation}
P_{\{J_{ij}\}}(q)=\sum_{S,S'}W_S W_S' \delta (q-q_{SS'}),
\label{eq:overlap-distribution}
\end{equation}
where $W_S$ is the probability of finding the system in macrostate $S$. 

In general, the macrostates can depend on the specific choice of the 
$\{J_{ij}\}$ but for the SK model  the average of $P_{\{J_{ij}\}}(q)$ 
can be calculated, as also other more 
complicated distributions of the $q_{SS'}$, such as the correlation 
of pairwise overlap distributions for 3 macrostates $S,S',S''$.

For a simple (non-complex) system there is only one thermodynamically relevant macrostate and hence 
$\overline{P(q)}$ has a single delta function peak; at $q=0$ for a paramagnet (in the absence of an external 
field) and at $q=m^2$ for a ferromagnet, where $m$ is the magnetization per spin. 
In contrast, in a complex system $\overline{P(q)}$ has structure indicating many relevant 
macrostates\footnote{The overline indicates an average over the quenched disorder.}. This is the case for the 
SK model beneath a critical temperature and for sufficient frustration, as measured by the ratio of the 
standard deviation of $\mathsf{P}(J)$ compared with its mean. Furthemore other measures of the $q$-distribution 
indicate a hierarchical structure, ultrametricity and a phylogenic-tree structure for relating overlaps of 
macrostates, chaotic evolution with variations of global parameters, and also non-self-averaging of appropriate 
measures.

These observations and others give substance to and quantify  the rugged landscape 
picture with macrostate barriers impenetrable on timescales becoming infinite 
with $N$. For finite-ranged spin glasses this picture must be relaxed  
to have only finite barriers, but still with a non-trivial phase transition to a glassy state.

The macroscopic dynamics in the spin glass phase also shows novel and interesting glassy 
behaviour\footnote{There are several possible microscopic dynamics that leads to the same equilibrium/Gibbsian 
measure, but all such employing local dynamics lead to glassiness.}, never equilibrating and having significant 
deviations from the usual fluctuation-dissipation relationship\footnote{ Instead one finds a modified 
fluctuation-dissipation relation with the temperature normalized by the instantaneous auto-correlation.}. 

A brief introduction to the methodolgies to arrive at these conclusions is deferred to a later section. 

\section{Transfers and Extensions}

The knowledge gained from such spin glass studies has been applied to increasing understanding of several other 
physically different systems and problems, via mathematical and conceptual transfers and extensions. Conversely 
these other systems have presented interesting new  challenges for statistical physics. In this section we 
shall illustrate this briefly via discussion of some of these transfers and stimulating extensions.

In static/thermodynamic extensions there exist several different analogues of the quenched and annealed 
microscopic variables, $\{J\}$ and $\{\sigma\}$ above, and of the intensive controls, such as $T$. Naturally, 
in dynamics of systems with quenched disorder the annealed variables (such as the $\{\sigma\}$ above) become 
dynamical,  but also one can consider cases in which the previously quenched parameters are also dynamical but 
with slower fundamental microscopic timescales\footnote{Sometimes one speaks of fast and slow microscopic 
variables but it should be emphasised that these refer to the underlying microscopic time-scales. Glassiness 
leads to much slower macroscopic timescales.}. 

\subsection{Optimization and satisfiability}
Already in section (\ref{sec:SKDean}),  one example of eqn.(\ref{eq:SK}) as an optimization problem was given 
(the Dean's Problem). Another classic hard computer science optimization problem is that of equipartitioning a 
random graph so as to minimise the cross-links. In this case the cost function to minimise can be again be 
written as in eqn. (\ref{eq:SK}), now with the  $\{i\}$ labelling vertices of the graph, the $\{J_{ij}\}$ equal 
to $1$ on edges/graph-links between vertices and zero where there is no link between $i$ and $j$, the 
$\{\sigma_{i}={\pm}\}$ indicating whether vertices ${i}$ are in the first or second partition and with the 
frustrating constraint $\sum_{i}{\sigma_{i}}=0$ imposing equipartitioning. Without the global constraint this 
is a random ferromagnet, but with it the system is in the same complexity class as a spin glass. 

Another classic hard optimization problem that extends eqn. (\ref{eq:SK}) in an apparently simple way but in 
fact leads to new consequence is that of random $K$-satisfiability ($K$-SAT) \cite{KS}. Here the object is to 
investigate the simultaneous satisfiability of many,  $M$, randomly chosen clauses, each  made up of $K$ 
possible microscopic conditions involving a large number, $N$, of binary variables. Labelling the variables 
$\{\sigma_{i}\}=\{\pm{1}\}$ 
and writing $x_{i}$ to indicate $\sigma_{i}=1$ and $\overline{x_{i}}$ to indicate $\sigma_{i}=-1$, a 
$K$-clause has the form 
\begin{equation}
(y_{i_{1}} {\rm{or}} \; y_{i_{2}} {\rm{or ....}} y_{i_{K}}); \; i = 1,..M
\label{eq:K-clause}
\end{equation}
where the $y_{i_{j}}$ are 
 $x_{i_{j}}$ or $\overline{x_{i_{j}}}$. In Random K-Sat the $\{i_{j}\}$ are chosen randomly from the $N$ 
possibilities and the choice of $y_{i_{j}} = x_{i_{j}} \; {\rm{or}} \; \overline{x_{i_{j}}}$ is also random, in 
both cases then quenched. In this case one finds, for the thermodynamically relevant typical system,  that 
there are two transitions as the ratio $\alpha=M/N$ is increased in the limit $N \to \infty$; for $\alpha > 
\alpha_{c1}$ it is not possible to satify all the clauses simultaneously (UNSAT), for $\alpha < \alpha_{c1}$ 
the problem is satisfiable in principle (SAT), but for $\alpha_{c2} < \alpha < \alpha_{c1}$ it is very 
difficult to satisfy (in the sense that all simple local variational algorithms stick) and this region is known 
as HARD-SAT. These distinctions are attributable to regions of fundamentally different fractionation of the 
space of satisfiability, different levels of complexity.

\subsection{K-spin glass}
In fact, again there was a stimulating precursor of this $K$-SAT discovery in a ``{\it{what-if}}'' extension of 
the SK model \cite{Crisanti,CSH} in which the 2-spin interactions of eqn. (\ref{eq:SK}) are replaced by K-spin 
interactions:
\begin{equation}
H_{K}=-\sum_{(i_{1},i_{2}...i_{K})} J_{i_{1},i_{2}...i_{K}} \sigma_{i_{1}}\sigma_{i_{2}}....\sigma{i_{K}}
\label{eq:Kspin}
\end{equation}
in which the $J_{i_{1},i_{2}...i_{K}}$ are again chosen randomly and independently from an intensive 
distribution of zero mean. In this case, two different phase transitions are observed as a function of 
temperature,  a lower thermodynamic transition and a dynamical transition that is at a slightly higher 
temperature, both to complex spin glass phases. The thermodynamic transition represents what is achievable in 
principle in a situation in which all microstates can be accessed; the dynamical transition represents the 
situation where the system gets stuck and cannot explore all the  possibilities, analogues of HARDSAT-UNSAT and 
SAT-HARDSAT. 

The $K$-spin glass is also complex with a non-trivial overlap distribution function $P(q)$ but now the state 
first reached as the transitions are crossed has a different structure from that found for the 2-spin case. Now
\begin{equation}
\overline{P(q)} = (1-x) \delta({q - q_{min}}) + x\delta({q-q_{\rm{max}}});
\label{eq:KspinP(q)}
\end{equation}
in contrast with the SK case where there is continuous weight below the maximum $q_{max}$. The two delta 
functions demonstrate that there is still the complexity of many equivalent but different macrostates, but now 
with equal mutual orthogonalities (as compared with the 2-spin SK case where there is a continuous range of 
macrostates). This situation turns out to be quite common in many extensions beyond SK.

\subsection {Statics, dynamics and temperature}
At this point it is perhaps useful to say a few more words about the differences between statics/thermodynamics 
and dynamics in statistical physics, and about types of micro-dynamics and analogues of temperature. 

In a physical system one often wishes to study thermodynamic equilibrium, assuming all microstates are 
attainable if one waits long enough. In optimization problems one typically has two types of problem; the first 
determining what is attainable in principle, the second considering how to attain it. The former is the 
analogue of thermodynamic equilibrium, the latter of dynamics. 

In a physical system the true microscopic dynamics is given by nature. However, in optimization studies the 
investigator has the opportunity to determine the micro-dynamics through the computer algorithms he or she 
chooses to employ .

Temperature enters the statistical mechanics of a physical problem in the standard Boltzmann-Gibbs ensemble 
fashion, or as a measure of the stochastic noise in the dynamics. We have already noted that it can also enter 
an optimization problem in a very similar fashion if there is inbuilt uncertainty in the quantity to be 
optimized. But stochadtic noise can also usefully  be introduced into the artificial computer algorithmic 
dynamics used to try to find that optimum. This is the basis of the optimization technique of simulated 
annealing where noise of  variance $T_{A}$ is deliberately introduced to enable the probabalistic scaling of 
barriers, and then gradually reduced to zero \cite{KGV}.

\subsection{Neural networks}

The brain is made up of a very large number of neurons, firing at different rates and extents,
interconnected by an even much larger number of synapses, both excitory and inhibitory. In a simple model due 
to 
Hopfield \cite{Hopfield}
one can consider a cartoon describable again by a control function of the form of eqn. (\ref{eq:SK}).
In this model the neurons $\{i\}$ are idealized by binary McCullough-Pitts variables $\{\sigma_{i} =\pm{1}\}$, 
the synapses by $\{J_{ij}\}$, positive for excitatory and negative for inhibitory,
with stochastic neural microdynamics of effective temperature $T_{\rm{neural}}$ emulating the width of the 
sigmoidal 
response of a neuron's output to the combined input from all its afferent synapses, weighted by the 
corresponding 
activity of the afferent neurons. 

The synapses are distributed over both signs, yielding frustration and apparently random at first sight. 
However actually  they are coded to enable attractor basins related to memorized patterns of the neural 
microstates 
$\{\xi_{i}^{\mu}\}; \; \mu=1,...p= \alpha N$. The similarity of a neural micro-state to a pattern $\mu$ is 
given by an overlap
\begin{equation}
m^{\mu}=N^{-1}\sum_{i} \langle \sigma_i \rangle_{S} \xi_{i}^{\mu}.
\label{eq:neural overlap}
\end{equation}
Retrieval of memory $\mu$ is the attractor process in which a system started with a 
small $m^\mu$ iterates towards a large value of $m^\mu$. 

In Hopfield's original model he took the $\{J_{ij}\}$ to be given by the Hebb-inspired form
\begin{equation}
J_{ij}=p^{-1}\sum_{\mu}\xi^{\mu}_{i} \xi^{\mu}_{j}
\label{eq:Hopfield}
\end{equation}
with randomly quenched $\{\xi_{i}\}$ \footnote{{\it{i.e.}} uncorrelated patterns}.
For $\alpha$ less than a $T_{neural}$-dependent critical value $\alpha_c(T_{neural})$ patterns can be 
retrieved. Beyond it only quasi-random spin glass minima unrelated to the memorised patterns remain ( and still 
only for $T_{neural}$ not too large). However, other $\{J_{ij}\}$ permit 
a slightly larger capacity (as also can occur for correlated patterns).

Again the landscape cartoon is illustratively useful. It can be envisaged as one for $H_{\{{J_{ij}\}}}$ as a 
function of the neural microstates (of all the neurons), with the dynamics one of motion in that landscape, 
searching for minima using local deviation attempts. The memory basins are large minima. Clearly one would like 
to have many different retrievable memories. Hence frustration is necessary. But equally, too much frustration 
would lead to a spin-glass-like state with minima unrelated to learned memories. 

This cartoon also leads immediately to the recognition that learning involves modifying the landscape so as to 
place the attractor minima around the states to be retrieved. This  extension  can be modelled minimally via a 
system of coupled dynamics of neurons whose state dynamics is fast (attempting retrieval or generalization)  
and synapses that also vary dynamically but on a much slower timescale and in response to external 
perturbations (yielding learning) \cite{Penney93}.

\subsection{Minority game}

More examples of many-body systems with complex macrobehaviour are to be found in social systems,
in which the microscopic units are people (or groups of people or institutions), sometimes co-operating, 
often competing. Here explicit discussion will be restricted to one simple model, the Minority Game 
\cite{Challet}, 
devised to emulate some features
of a stockmarket. $N$ `agents' play a game 
in which at each time-step each agent makes one of two choices 
with the objective to make the choice which is in the minority\footnote{The philosophy is that one 
gets the best price by selling when most want to buy or 
buying when most want to sell.}. They have no direct knowledge of one
another but  (in the original version) make their choices based on the commonly-available knowledge of
the historical actual minority choices, using their own individual stategies and experience to 
make their own decisions.
In the spirit of minimalism we consider all agents (i) to have the same `memories', of the minority choices for
the last $m$ time-steps, (ii) to each have two strategies given by randomly 
chosen and quenched Boolean operators
that, acting on the $m$-string of binary entries representing the minority choices for the 
last $m$ steps, output a  binary instruction on the choice to make, (iii) using a personal `point-score' to
keep tally of how their strategies would have performed if used, increasing the score each time they would
have chosen the actual minority, and (iv) using their strategy with the larger point-score.  
Frustration is represented in the minority requirement,  while quenched 
disorder arises in the random choice of individual strategies.
  
Simulational studies of the `volatility', the standard deviation 
of the actual minority choice, shows (i) a deviation from individually random choices,
indicating correlation through the common information, (ii) a cusp-minimum at a critical value 
$\alpha_c$ of the ratio of the 
information dimension to the number of agents $\alpha = D/N = 2^{m}/N$, 
suggesting a phase transition at $\alpha_c$, (iii) ergodicity for $\alpha > \alpha_c$
but non-ergodic dependence on the point-score initialization for  $\alpha < \alpha_c$, indicating that the 
transition 
represents the onset of complexity.
This is reminiscent of the cusp and the ergodic-nonergodic transition observed in the susceptibilities
of spin glass systems
as the temperature is reduced through the spin glass transition.

Furthermore, this behaviour is essentially unaltered if the `true' history is replaced by a 
fictitious `random' history at each step, with all agents being given  the same false history, indicating that 
it principally
represents a carrier for an effective interaction between the agents. Indeed, generalising to 
a $D$-dimensional random history information  space, considering this as a vector-space and 
 the strategies as 
quenched $D$-vectors of components$\{R_{i}^{s,\mu}\}; \; s=1,2, \; \mu=1,..D$ in that space, and 
averaging over the stochastically random `information',  one is led to
an effective control function analagous to those of eqn. (\ref{eq:SK}) and eqn. (\ref{eq:Hopfield}) with $p$ 
replaced by $\alpha$, now  $\{\xi_{i} =(R_{i}^1 - R_{i}^2)/2\}$, an extra multiplicative minus sign on the 
right hand side  
of eqn. (\ref{eq:Hopfield}), and also a random-field term dependent upon  the $\{\xi_{i}\}$ and $\{\omega_{i} 
=(R_{i}^1 + R_{i}^2)/2\}$. As noted, there is an ergodic-nonergodic transition at a crtitical $\alpha$,  but 
now the picture is one of the $\{\xi_{i}\}$ as repellers rather than the attractors of the Hopfield 
model\footnote{One can make the model even more minimal by allowing each agent only one strategy $\{\xi_{i}\}$ 
which (s)he either follows if its point-score is positive or acts oppositely to if the point-score is negative. 
This removes the random-field term and also the cusp in the {\it{tabula rasa}} volatility, but retains the 
ergodic-nonergodic transition\cite{Galla}.}.

The typical behaviour of this system, as for the spin glasses,  can be studied using a dynamical generating 
functional method
\cite{Coolen}, averaged over the choice of quenched strategies, in a manner outlined below. The averaged 
many-body system  can then be mapped into an effective {\it{representative agent ensemble}} with {\it{memory}} 
and {\it{coloured noise}}, with both  the noise correlations and the memory kernel determined self-consistently 
over the ensemble. Note that this is in contrast to (and corrects) the common assumption of a single 
deterministic representative agent. The phase transition from ergodic to non-ergodic is manifest by a 
singularity in the two-time point-sign correlation function 
\begin{equation}
C(t,t')=N^{-1}\sum_{i} \overline{{\rm{sgn} (p_{i}(t) \rm{sgn} (p_{i}(t')}}
=\langle {\rm{sgn} (p(t) \rm{sgn} (p(t')} \rangle_{ens}
\label{eq:corrn_fn}
\end{equation}
where the first equality refers to the many-body problem and the second its equivalence in the effective agent 
ensemble.

\section{Methodologies}

For systems in equilibrium, physical observables are given by $\ln Z$ evaluated for the specific instance of 
any quenched parameters, or strictly the generalized generating function $\ln Z(\{\lambda\})$ where the $\{ 
\lambda \}$ are generating fields to be taken to zero after an appropriate operation (such as 
$\partial/\partial\lambda$) is performed. Hence the average over quenched disorder is given by 
$\overline{\ln Z }$. One would like to perform the average over quenched disorder explicitly to yield an 
effective system. However, since $Z$ is a sum over exponentials of a function of the variables,  $\ln Z$ is 
difficult to average directly so instead one uses the relation
\begin{equation}
\ln Z = Lim_{n \to 0}\;  n^{-1} (Z^n -1)
\label{eq:Replica_trick}
\end{equation}
and interprets the $Z^n$ as corresponding to a system whose variables have   extra `replica' labels, $\alpha = 
1,..n$,  for which one can then  average the partition function, an easier operation, at the price of needing 
to take the eventual limit $n \to 0$. The relevant `order parameters' are  then correlations between replicas
\begin{equation}
q^{\alpha \beta} = N^{-1}\sum_{i}\langle \sigma_{i}^\alpha \sigma_{i}^ \beta \rangle_T
\label{eq:EA_OP}
\end{equation}
where $\langle..\rangle_T$ refers to a thermal average in the effective post-averaging system. This order 
parameter is non-zero in the presence of frozen order, but more interestingly (and subtly)  also exhibits the 
further remarkable feature of spontaneous replica symmetry breaking, indicating complexity. After further 
subtleties beyond the scope of this
short introduction, there emerges an order function $q(x); x \in [0.1]$ from which the average overlap function 
is obtained by
\begin{equation}
\overline{P(q)} = dx/dq
\label{eq:P(q)}
\end{equation}

For dynamics the analogue of the partition function $Z$ is a generating functional, which may be written 
symbolically as
\begin{equation}
Z_{dyn}=\int  \prod_{\rm{all \; variables, \; all\; times}}
 {\delta (\rm{microscopic \;  eqns. \; of \; motion})} \exp(\{\lambda \phi\})
\label{eq:gen_fnl}
\end{equation}
where the $\phi$ symbolize the microscopic variables and a Jacobian is implicit. Averaging over the quenched 
disorder now induces interaction between epochs and integrating out the microscopic variables results in the 
effective single agent ensemble formulation, as well as emergent correlation and response functions as the 
dynamic order parameter analogues of the static inter-replica overlaps, exhibiting non-analyticity at a phase 
transition to non-ergodicity.

\section{Conclusion}

A brief illustration has been presented of how complex co-operative behaviour arises in many body systems due 
to the combination of frustration and disorder in the microscopics of even very simply formulated problems with 
very few parameters. Such systems are not only examples of Anderson's famous quotation
{\it{``More is different''}} but also demonstrate that {\it{frustration and disorder in microscopics can lead 
to complexity 
in macroscopics}}; {\it{i.e.}} many and complexly related {\it{different}}s. Furthermore, this complexity 
arises in\
systems with very simple few-valued microscopic parameters; {\it{complexity is {\textbf {not}} the same as 
{complication}}}
 and does not require it. 

There has also been  demonstrated  valuable transfers  between systems that appear very different at first 
sight, through the media of mathematical modelling, conceptualization and investigatory methodologies, a 
situation reminiscent of the successful use of the Rosetta stone in learning an unknown language script by 
comparison with another that carries the same message in a different format. 

The perspective taken has been of statistical physics, but it must be emphasised that the stimulation has been 
both from and to physics, since many of these  complex systems  have interesting features in their microscopic 
underpinning that are richer than those in the physics of conventional dictionary definition and provide new 
challenges to the  physicist.

Also of note is how a {\it{blue skies}} attempt to understand some obscure magnetic alloys through soluble but, 
for the experimental alloys, unphysical modelling has led to an explosion of appreciation of new concepts, 
understanding and application of ideas and methologies throughout an extremely wide range of the sciences.

\section{Acknowledgements}

The author thanks the Leverhulme Trust for the current award of an Emeritus Fellowship and  the UK EPSRC, the 
EU and
the ESF for earlier support over many years during the development of the work reported here. 
He also thanks  many colleagues
throughout the world for collaborations and valuable discussions;
most of their names are given  in the last slide of his 2010 Blaise Pascal lecture that can be found at 
\cite{BP}.

\end{document}